\documentclass[aps,prl,twocolumn,showpacs,amsmath,amssymb]{revtex4}
\usepackage{epsfig}
\usepackage{bm}
\def\be{\begin{equation}}
\def\ee{\end{equation}}

\begin{document}
\title{Two-particle Aharonov-Bohm effect and Entanglement in the
electronic Hanbury Brown Twiss set-up} 
\author{P. Samuelsson,
E. V. Sukhorukov and M. B\"uttiker} \affiliation{D\'epartement de
Physique Th\'eorique, Universit\'e de Gen\`eve, CH-1211 Gen\`eve 4,
Switzerland} \date{\today}

\begin{abstract}
We analyze a Hanbury Brown Twiss geometry in which particles are
injected from two independent sources into a mesoscopic conductor in
the quantum Hall regime. All partial waves end in different reservoirs
without generating any single particle interference, in particular,
there is no single particle Aharonov-Bohm effect. However, exchange
effects lead to two-particle Aharonov-Bohm oscillations in the
zero-frequency current cross-correlations. We demonstrate that this is
related to two-particle orbital entanglement, detected via violation
of a Bell Inequality. The transport is along edge states and only
adiabatic quantum point contacts and normal reservoirs are employed.
\end{abstract}

\pacs{73.23.-b, 05.40.-a, 72.70.+m, 74.40.+k} 
\maketitle 

Intensity correlations of photons became of interest with the
invention by Hanbury Brown and Twiss (HBT) of an interferometer which
permitted them to determine the angular diameter of visual stars
\cite{hbt}. The HBT effect contains two important distinct but
fundamentally interrelated effects: First, light from different,
completely uncorrelated, portions of the star gives rise to an
interference effect which is visible in intensity correlations but not
in the intensities themselves. This is a property of two particle
exchange amplitudes. Exchange amplitudes are a quantum mechanical
consequence of the indistinguishability of identical
particles. Second, there is a direct statistical effect since photons
bunch whereas fermions anti-bunch. Fundamentally both of these effects
are related to the symmetry of the multiparticle wave function under
exchange of two particles. For photons emitted by a thermal source a
classical wave field explanation of the HBT-effect is possible. A
quantum theory was put forth by Purcell\ \cite{pruc}. For fermions, no
classical wave theory is possible.

It has long been a dream to realize the electronic equivalent of the
optical HBT experiment. This is difficult to achieve with field
emission of electrons into vacuum because the effect is quadratic in
the occupation numbers. This difficulty is absent in electrical
conductors where at low temperatures a Fermi gas is completely
degenerate. Experiments demonstrating fermionic anti-bunching in
electrical conductors were reported by Oliver et al.\ \cite{oliv},
Henny et al.\ \cite{henny} and Oberholzer et al.\ \cite{ober}. Only very
recently was a first experiment with a field emission source
successful\ \cite{vacuum}. In contrast, to date, there is no
experimental demonstration of two-electron interference.

In electrical conductors "beams" can be realized in high-magnetic
fields in the form of edge states\ \cite{halp}. Edge channels permit
the transport of electrons over (electronically) large distances. In
the quantized Hall state\ \cite{vK} scattering out of an edge state is
suppressed\ \cite{mb88}. The second element needed to mimic optical
geometries, the half-silvered mirror, is similarly available in the
form of quantum point contacts \cite{qpc,noiseqpc} (QPC's). Indeed in
high magnetic fields a QPC permits the separate measurement of
transmitted and reflected carriers \cite{mb90}. A Mach-Zehnder
interferometer with edge states was recently realized \cite{ji}. This
shows that it is possible to implement arrangements of linear optics
\cite{knill} in electrical conductors.

Here we propose an implementation of the HBT-experiment in an
electrical conductor in the quantum Hall regime. In the set-up, there
is no single particle interference, however, two-particle interference
is manifested as a magnetic flux dependence of the current
correlators, a two-particle Aharanov-Bohm effect. We show that this
two-particle effect is closely related to orbital entanglement\
\cite{sam1} of electron-hole pairs, recently proposed by Beenakker et
al.\ \cite{been}, as well as of pairs of electrons. The entanglement is
detected via a violation of a Bell Inequality. Only normal electronic
reservoirs, adiabatic QPC's and zero-frequency correlators are
employed, greatly simplifying an experimental realization.

\begin{figure}[t]
\centerline{\psfig{figure=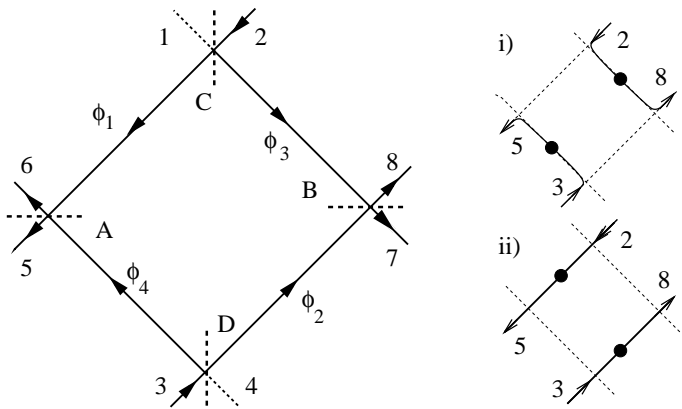,width=9.0cm}}
\caption{Left: Two-source, four-detector optical Hanbury Brown Twiss
geometry. Two beams, incident from $2$ and $3$ are split at the
mirrors $C$ and $D$ and impinge on the mirrors $A$ and $B$ to reach
detectors along the directions $5$ to $8$. Right: The two scattering
processes i) and ii) contributing to the phase dependence of the
correlators (see text).}
\label{fig1}
\end{figure}

\begin{figure}[t]
\centerline{\psfig{figure=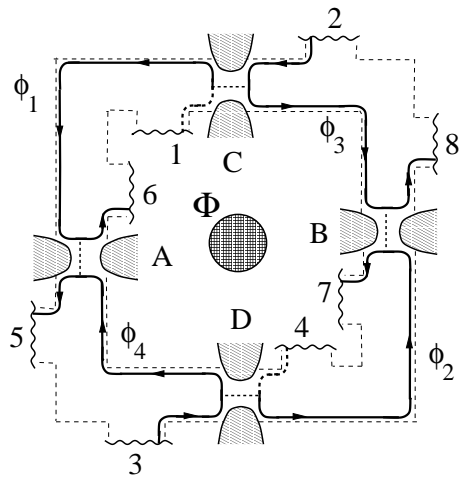,width=6.0cm}}
\caption{Two-source, four detector electrical Hanbury Brown Twiss
geometry: a rectangular Hall bar with inner and outer edges (thin
dashed lines) and four quantum point contacts (grey shaded). Contacts
$2$ and $3$ are sources of electrons (a voltage $eV$ is applied
against all other contacts which are at ground). Electrons follow edge
states (thick black lines) in the direction indicated by the arrows
and pick up phases $\phi_1$ to $\phi_4$. An Aharanov-Bohm flux $\Phi$
penetrates the center of the sample (shaded) but has no influence on
the single particle properties. However, the current correlations are
essentially dependent on the phases $\phi_i$ and the Aharonov-Bohm
flux. We note that not all contacts have to be realized experimentally
to investigate all physical phenomena discussed.}
\label{fig2}
\end{figure}

An optical configuration with two independent sources\ \cite{yust} is
shown in Fig.\ \ref{fig1}. It is a table top equivalent to the stellar
interferometer experiment of HBT. Fig.\ \ref{fig2} shows the
implementation of this configuration in an electrical conductor. The
geometry has no interfering orbits. Electron waves incident at the
$i-th$ QPC (with $i =A$ to $D$) are transmitted with amplitude
$\sqrt{T_{i}}$ and reflected with amplitude $\sqrt{R_{i}}$ with $T_{i}
+ R_{i} = 1$. Along the edge states electrons accumulate phases 
$\phi_1$ to $\phi_4$. The overall scattering
behavior is determined by the global scattering matrix
$s_{\alpha\gamma}$ which gives the current amplitude at contact
$\alpha$ in terms of the current amplitude at the incident contact
$\gamma$. Since there are no interfering orbits, the global scattering
matrix elements depend only in a trivial way on the phases.  A
particle leaving source contact $2$ can after transmission through QPC
$C$ reach either contact $5$ or $6$. For instance, the scattering
matrix element $s_{52} = \sqrt{T_{A}} \exp(i\phi_1) \sqrt{T_{C}}$ and
similarly for all other elements of the s-matrix.  Since the
conductance matrix elements are determined by transmission
probabilities, it follows immediately that in the set-up of
Fig.\ \ref{fig2} all conductance matrix elements are
phase-insensitive. For instance a voltage applied at contact $2$
generates a current at contact $5$, giving rise to a conductance
$G_{52} = -(e^{2}/h)T_{A} T_{C}$.  The conductances are thus determined
only by products of transmission and reflection probabilities of the
QPC's.

Let us now evaluate the current-current correlations for the geometry
of Fig.\ \ref{fig2}. The zero-frequency cross-correlations
$S_{\alpha\beta}$ of the current fluctuations $\Delta I_{\alpha}$ and
$\Delta I_{\beta}$ are defined through
\begin{equation}
S_{\alpha\beta} = 
\int dt \langle \left[\Delta I_{\alpha}(t)\Delta I_{\beta}(0)+\Delta
I_{\beta}(0) \Delta I_{\alpha}(t)\right] \rangle.
\end{equation}
Containing two current operators, the correlator provides information about the two-particle 
properties of the system. Following the scattering approach to noise correlators in Ref.\ \cite{mb90},
 the expression for the cross-correlations (at contact $\alpha\in 5,6$ and $\beta \in 7,8$) is given in terms of the scattering amplitudes  as
\begin{equation}
S_{\alpha\beta}=-(2e^2/h)\int dE |s_{\alpha2}^*s_{\beta2}+s_{\alpha3}^*s_{\beta3}|^2 (f-f_0)^2 
\label{crosscorr}
\end{equation}
where $f$ is the Fermi distribution function of reservoirs $2$ and $3$
(at a voltage bias $eV$) and $f_0$ the distribution function of the
other reservoirs (grounded). A corresponding calculation of the
light-intensity cross-correlations in the optical HBT geometry in
Fig.\ \ref{fig1} with thermal sources, would give the same result as
in Eq.\ (\ref{crosscorr}) but with opposite sign, an effect of
changing from fermionic to bosonic statistics of the carriers. 

The basic scattering processes contributing to the correlator are
clearly illustrated by considering the simplest case with transmission
and reflection probabilities of all QPC's equal to $1/2$. The
correlation function of the currents at e.g. contact $5$ and $8$ is
then, at zero temperature
\begin{equation}
S_{58} = - (e^2/4h) |eV| \left[1 + \cos (\phi_1 + \phi_2 - \phi_3 -
\phi_4)\right],
\label{hbtphases}
\end{equation}
depending in an essential way on the phases $\phi_1$ to $\phi_4$ in
Fig.\ \ref{fig2}. The phase dependent term results from the process
where one particle is emitted from contact 2 and one from contact
3. Detecting one particle in $5$ and one in $8$, we can quantum
mechanically not distinguish which paths the individual particles
took, i) from $2$ to $8$ and from $3$ to $5$ or ii) from $2$ to $5$
and from $3$ to $8$ [See Fig. \ref{fig1}]. As a consequence, the
amplitudes $\mbox{exp}[i(\phi_1+\phi_2)]$ and
$\mbox{exp}[i(\phi_3+\phi_4)]$ of the respective processes i) and ii)
must be added, giving rise to the interference term $\cos (\phi_1 +
\phi_2 - \phi_3 - \phi_4)$ in Eq. (\ref{hbtphases}). We emphasize that
it is the fact that both sources 2 and 3 are active that gives rise to
this phase dependence. The phase independent term is the sum of the
correlations that are obtained if only source $2$ is active and if
only source $3$ is active. 

In an experiment, it is possible to modulate the phases with the help
of gates which lengthen or shorten the paths of the edge states in
Fig. \ref{fig2}. The phases can as well be modulated with the help of
an Aharonov-Bohm (AB)-flux \cite{ab} through the center of the
structure. While there are no single particles trajectories which
coherently enclose the flux, the paths of two particles, one emitted
by $2$ and the other emitted by $3$, have the possibility to enclose
the flux. The flux contributes a positive phase to $\phi_1$ and
$\phi_2$ and a negative phase to $\phi_3$ and $\phi_4$ to give a total
additional phase contribution of $2\pi \Phi/\Phi_{0}=\oint d{\bf
l}\cdot{\bf A}$, where ${\bf A}$ is the vector potential and
$\Phi_0=h/e$ the single charge flux quantum. The total phase in
Eq. (\ref{hbtphases}) is then $\phi_1 + \phi_2 - \phi_3 - \phi_4 + 2
\pi \Phi/\Phi_{0}$. The possibility of such an AB-effect
due to two-particle exchange was recognized in early work on noise
\cite{mb91,prl92} and in the co-tunneling current \cite{dles}. The
geometry of Fig. 2 is unique in that the conductances (second order
interference) exhibit no AB-effect but current correlators (fourth
order interference) are sensitive to the variation of an AB-flux.

Apart from this two-particle AB-effect, is the opposite sign of the
correlator in Eq. (\ref{crosscorr}), resulting from the different
statistics of the carriers, the only significant difference between
the electronic and the photonic HBT-setup? The answer is no. As we now
show, due to the degeneracy of the electron sources at low
temperatures, for strongly asymmetric source QPC's $C$ and $D$ in
Fig. \ref{fig2}, an orbitally entangled \cite{sam1} electron-hole pair
\cite{been} state is emitted from $C$ and $D$. This has no counterpart
in the optical HBT-setup with thermal sources \cite{opticscom}.

We take the transmission and reflection probabilities at the QPC $C$
to be $T_C=1-R_C=T$ and at $D$ to be $T_D=1-R_D=R$ (scattering
probabilities at $A$ and $B$ are specified below). The many-body transport
state generated by the two independent sources is in second
quantization (suppressing the spin index) $|\Psi
\rangle=\prod_{0<E<eV}c_2^{\dagger}(E)c_3^{\dagger}(E)|0\rangle$,
where $|0\rangle$ is the ground state, a filled Fermi sea in all
reservoirs at energies $E<0$. The operator $c_{\gamma}^{\dagger}(E)$
creates an injected electron from reservoir $\gamma$ at energy
$E$. After scattering at $C$ and $D$, the state $|\Psi \rangle$
consists of two contributions in which the two particles fly off one
to $A$ and one to $B$, and of two contributions in which the two
particles fly both off towards the same detector QPC.

Consider now the case of strong asymmetry $R \ll 1$, where almost no
electrons are passing through the source QPC's towards $B$. We can
write the full state $|\Psi \rangle$ to leading order in $\sqrt{R}$ as
$|\Psi\rangle=|\bar 0\rangle+\sqrt{R}|\tilde \Psi\rangle$, with
\begin{equation}
|\tilde \Psi \rangle=\int_0^{eV}\hspace{-0.3cm}dE
\left[c_{3B}^{\dagger}(E)c_{3A}(E)-c_{2B}^{\dagger}(E)c_{2A}(E)\right]|\bar 0\rangle
\label{newgs}
\end{equation}
The second index of the electron operators, $A$ or $B$, denotes
towards which detector the particle is propagating.  Here we have
redefined the vacuum to be the completely filled stream of electrons,
$|\bar 0
\rangle=\prod_{0<E<eV}c_{2A}^{\dagger}(E)c_{3A}^{\dagger}(E)|0\rangle$. The
operators $c_{3A}(E)$ and $c_{2A}(E)$ describe hole excitations,
i.e. the removal of an electron at energy $E$ from the filled
stream. Due to the redefinition of the vacuum \cite{sam1}, we can
interpret the resulting state $|\tilde \Psi \rangle$ as describing a
superposition of "wavepacket"-like electron-hole pair excitations out
of the new vacuum, i.e. an orbitally entangled pair of electron-hole
excitations. This is equivalent to the recent findings by Beenakker et
al. \cite{been}, who discussed the generation of entangled
electron-hole pairs at a single non-adiabatic QPC. The state is
similar to the two-electron state emitted from a superconductor
contacted at two different points in space \cite{sam1,samcom}. The new
vacuum $|\bar 0 \rangle$ is noiseless and does not contribute to the
cross-correlators, even so it carries a current from the sources
to the detectors. The cross-correlation measurement is thus sensitive
only to $|\tilde \Psi\rangle$, the entangled electron-hole pair state.

Following our earlier work \cite{sam1,BIs}, this two-particle orbital
entanglement can be detected via violation of a Bell Inequality (BI)
\cite{bell}. In the limit $R\ll 1$, the time between emission of
successive electron-hole pairs $h/(eVR)$ is much larger than the
coherence time $\tau_C=h/eV$ of each pair. As a consequence, the
zero-frequency noise measurement works as a coincidence measurement
running over a long time. The BI can then be expressed directly in
terms of the zero-frequency cross-correlators.

Interestingly, the electron-hole entanglement is not the only feature
of the electric HBT-setup which has no counterpart in optics
\cite{opticscom}. The anti-bunching of electrons implies that no two
electrons can be emitted simultaneously from a single reservoir. As a
consequence, electrons emitted from a single source can not be
detected simultaneously in reservoirs $\alpha$ and $\beta$. Only the
process where one electron is emitted from $2$ and one from $3$ (shown
in Fig. \ref{fig1}), can lead to a joint detection in $\alpha$ and
$\beta$. As discussed above, since the paths of these two particles
can not be distinguished, the corresponding state is orbitally
entangled. The total state emitted from the contacts $C$ and $D$,
expressed in terms of electrons, is however a product state. Thus, the
process of jointly detecting one particle in $\alpha$ and one in
$\beta$ means effectively post-selecting \cite{Shih} a pair of
orbitally entangled electrons by the measurement.

This post-selected entanglement can be detected by a violation of a BI
formulated in terms of the joint detection probability of two
electrons. In optics, using photo-detectors, the joint probability of
detecting two photons is given by the theory of Glauber
\cite{Glauber}. In close analogy with Ref. \cite{Glauber} we define
the probability of simultaneous detection (at energy $0<E<eV$) of one
electron in detector $\alpha$ and one in $\beta$, as
\begin{eqnarray}
P_{\alpha\beta}\propto\langle
c^{\dagger}_{\beta}(t)c^{\dagger}_{\alpha}(t)c_{\alpha}(t)c_{\beta}(t)\rangle
\label{jdp}
\end{eqnarray}
where $c^{\dagger}_{\alpha}(t)=\int
dE~\mbox{exp}(iEt/\hbar)c_{\alpha}^{\dagger}(E)$. The probabilities
are normalized such that $\sum P_{\alpha\beta}=1$. In mesoscopic
systems, $P_{\alpha\beta}$ is difficult to measure directly. However,
as a non-local quantum mechanical correlator, it provides information
about the entanglement of two spatially separated particles in a
many-particle system.

For the setup in Fig. \ref{fig2}, $c_{\alpha}(t)$ and
$c^{\dagger}_{\beta}(t)$ anticommute. As a consequence
$P_{\alpha\beta}\propto \langle I_{\alpha}(t)I_{\beta}(t)\rangle$ and
we find
\begin{eqnarray}
P_{\alpha\beta}\propto S_{\alpha\beta}+2\tau_CI_{\alpha}I_{\beta}
\label{jdptot}
\end{eqnarray}
where $I_{\alpha}=(e^2/h)TV$ and $I_{\beta}=(e^2/h)RV$ are the
currents flowing into reservoirs $\alpha$ and $\beta$ and
$\tau_C=h/eV$ the coherence time. The zero-frequency correlator
$S_{\alpha\beta}$ is investigated by varying the transmission through
the two QPC's $A$ and $B$ which precede the detector reservoirs. This
is similar to schemes in optics where one varies the transmission to
the detectors with the help of polarizers. The transmission and
reflection probabilities through the detector QPC's are taken to be
$T_A=1-R_A=\sin^{2}\theta_A$ for $A$ and with $\theta_A \rightarrow
\theta_B$ for $B$. Then, Eq. (\ref{crosscorr}) gives
\begin{eqnarray}
S_{58}&=&-\frac{2e^2}{h}|eV|RT
\left[\sin^2\theta_A\sin^2\theta_B+\cos^2\theta_A\cos^2\theta_B
\right. \nonumber \\ 
&+&\left. 2\cos(\phi_0)\cos\theta_A\cos\theta_B\sin\theta_A\sin\theta_B\right]
\label{noise}
\end{eqnarray}
with $S_{67}=S_{58}$ and $S_{57}=S_{68}$ obtained from $S_{58}$ by
shifting $\theta_A \rightarrow \theta_A+\pi/2$. Here the phase
$\phi_0=\phi_1+\phi_2-\phi_3-\phi_4+2\pi\Phi/\Phi_0$. 

The BI, following Ref.\ \cite{clau}, is expressed in terms of
correlation functions
\begin{eqnarray}
&&E(\theta_A,\theta_B)=P_{58}+P_{67}-P_{57}-P_{68} \nonumber \\
 &&=\cos(2\theta_A)\cos (2\theta_B)+\cos(\phi_0)\sin
 (2\theta_A)\sin (2\theta_B).
\label{corrfunc2}
\end{eqnarray}
The BI is $-2\leq S_B \leq 2$, where the Bell parameter
$S_B=E(\theta_A,\theta_B)-E(\theta_A',\theta_B)+E(\theta_A',\theta_B)+E(\theta_A',\theta_B')$
with $\theta_A,\theta_A',\theta_B$ and $\theta_B'$ four different
measurement angles. Optimizing the angles, the maximum Bell parameter
is given by \cite{sam1} $S_B^{max}=2\sqrt{1+\cos^2(\phi_0)}$, i.e. the
BI can be violated for any $\cos(\phi_0)$. We note that for the
electron-hole pair entanglement discussed above, the same maximum Bell
parameter is obtained, for the BI expressed directly in terms of
zero-frequency correlators. Dephasing, due to e.g. a fluctuating
$AB$-phase or phases $\phi_i$, renormalizes $\cos (\phi_0) \rightarrow
\gamma \cos(\phi_0)$, eventually suppressing \cite{enav} the
entanglement for strong dephasing $\gamma \rightarrow 0$. However, the
BI can still be violated for arbitrary strong dephasing. We note that
$\gamma$ is just the visibility of the two-particle
AB-oscillations. This shows the strong connection between the orbital
entanglement and the two-particle AB-effect.

We emphasize that in contrast to the electronic HBT-setup, the BI can
not be violated in the setup in Fig. \ref{fig1} with thermal optical
sources. The reason for this is that the bunching of photons in the
thermal sources allows for two photons to be emitted simultaneously
from a single source. These additional two-photon scattering
processes, not present in the electronic case, are uncorrelated, i.e
not entangled. Formally, a calculation of the joint detection
probability \cite{Reid} gives the same result as in
Eq. (\ref{jdptot}). However, $S_{\alpha\beta}$ has opposite sign and
as a consequence, the correlation function in Eq. (\ref{corrfunc2})
should be multiplied with $1/3$, making a violation impossible.

We have treated only the case of integer quantum Hall states. The
fractional quantum Hall effect offers a wider, very interesting, area
for the examination of correlations \cite{safi} since in this case
fractional statistics is realized.

In conclusion, we have demonstrated the connection between the Hanbury
Brown Twiss effect, the two-particle Aharanov Bohm effect and orbital
entanglement. The simple adiabatic edge-state geometry described above
and the use of zero-frequency correlators brings experimental tests of
these effects within reach.

We acknowledge discussions with C.W.J. Beenakker, N. Gisin and
V. Scarani. This work was supported by the Swiss NSF and the program
for MANEP.


\begin{thebibliography}{02}

\bibitem{hbt} R. Hanbury Brown, and R.Q. Twiss, Nature {\bf 178}, 1046
(1956).


\bibitem{pruc}  E.M. Purcell, Nature {\bf 178}, 1449 (1956). For a
modern account, see G. Baym, arXiv:nucl-th/9804026.

\bibitem{oliv} W.D. Oliver {\it et al.}, Science {\bf 284}, 299 (1999).

\bibitem{henny} M. Henny, {\it et al.}, Science {\bf 284}, 296 (1999).

\bibitem{ober} S. Oberholzer {\it et al.}, Physica {\bf 6E}, 314 (2000). 


\bibitem{vacuum} H. Kiesel, A. Renz, and F. Hasselbach, Nature {\bf
418}, 392 (2002).

\bibitem{halp}  B.J. Halperin, Phys. Rev. B {\bf 25}, 2185 (1982). 

\bibitem{vK} K. von Klitzing, G. Dorda, and M. Pepper,
Phys. Rev. Lett. {\bf 45}, 494 (1980).
                
\bibitem{mb88}  M. B\"uttiker, Phys. Rev. B {\bf38}, 9375 (1988).                
\bibitem{qpc} B.J. van Wees {\it et al.}, Phys. Rev. Lett. {\bf 60},
848 (1988); D.A.  Wharam, {\it et al.}, J. Phys. C {\bf 21} L209
(1988). 

\bibitem{noiseqpc} M.I. Reznikov {\it et al.}, Phys. Rev. Lett. {\bf
75}, 3340 (1995); A. Kumar {\it et al.}, {\it ibid} {\bf 76}, 2778
(1996).

\bibitem{mb90} M. B\"uttiker, Phys. Rev. Lett. {\bf 65,} 2901 (1990);
Phys. Rev. B {\bf 46}, 12485 (1992).

\bibitem{ji} Ji, Y. {\it et al.}, Nature {\bf 422,} 415-418 (2003).

\bibitem{knill} E. Knill, R. Laflamme, and G.J. Milburn, Nature {\bf
409}, 46 (2001).

\bibitem{sam1}  P. Samuelsson, E.V. Sukhorukov, and M. B\"uttiker, Phys. Rev. Lett. (to appear) cond-mat/0303531.

\bibitem{been}  C.W.J. Beenakker {\it et al.} cond-mat/0305110.

\bibitem{yust}  B. Yurke, and D. Stoler, Phys. Rev. A {\bf 46}, 2229 (1992). 
  
\bibitem{ab} Y. Aharonov, and D. Bohm,  Phys. Rev. {\bf 115}, 485 (1959).      
       
\bibitem{mb91}  M. B\"{u}ttiker, Physica B {\bf 175}, 199 (1991).
               
\bibitem{prl92} M. B\"uttiker, Phys. Rev. Lett. {\bf 68}, 843 (1992).

\bibitem{dles} D. Loss, and E.V. Sukhorukov, Phys. Rev. Lett. {\bf
84}, 1035 (2000).

\bibitem{opticscom} 
In optics, systems with independent, {\it
non-thermal}, sources similar to Fig. \ref{fig1} have been used for
generation of entanglement, see e.g. B. Pittman, and
J.D. Franson, Phys. Rev. Lett. {\bf 90}, 240401 (2003) and Refs therein.

 
\bibitem{samcom} For the two-electron (not electron-hole) state
considered in Ref. \cite{sam1}, there is no two-particle AB-effect.

\bibitem{BIs} For alternative formulations of BI's, see X. Ma\^itre,
W. D. Oliver, Y. Yamamoto, Physica E {\bf 6} 301 (2000); S. Kawabata,
J. Phys. Soc. Jpn. {\bf 70}, 1210 (2001); N.M. Chtchelkatchev {\it et
al.}, Phys. Rev. B {\bf 66}, 161320 (2002); L. Faoro, F. Taddei,
R. Fazio, cond-mat/0306733.
               
\bibitem{bell}  J.S. Bell, Physics {\bf 1}, 195 (1965). 

\bibitem{Shih}
Y.H. Shih and C.O Alley, Phys. Rev. Lett. {\bf 26}, 2921 (1988).

\bibitem{Glauber}
R. J. Glauber, Phys. Rev. {\bf 130}, 2529 (1963).

\bibitem{clau}  J.F. Clauser {\it et al}, Phys. Rev. Lett. {\bf 23},
880 (1969); J.F Cluaser and M.A. Horne, Phys. Rev. D {\bf 10}, 526 (1974).

\bibitem{enav} For lengths of the edgestates
$L_{CA}+L_{DB}-L_{CB}-L_{DA} \ll \hbar
v_F/eV$, there is no suppression due to energy averaging.

\bibitem{Reid} See e.g. M.D. Reid and D.F. Walls, Phys. Rev. A, {\bf
34}, 1260 (1986).

\bibitem{safi} I. Safi, P. Devillard, and T. Martin,
Phys. Rev. Lett. {\bf 86,} 4628 (2001); S. Vishveshwara, cond-mat/0304568.

\end{thebibliography}
\end{document}